\documentclass[twocolumn,english,superscriptaddress,aps,pra,preprintnumbers,amsmath,amssymb,floatfix,groupedaddress]{revtex4-1}
\usepackage[T1]{fontenc}
\usepackage[utf8]{inputenc}
\usepackage{amsmath}
\usepackage{amssymb}
\usepackage{stmaryrd}
\usepackage{graphicx}
\usepackage{esint}
\usepackage{braket}
\usepackage{hyperref}

\makeatletter
\@ifundefined{textcolor}{}
{%
 \definecolor{BLACK}{gray}{0}
 \definecolor{WHITE}{gray}{1}
 \definecolor{RED}{rgb}{1,0,0}
 \definecolor{GREEN}{rgb}{0,1,0}
 \definecolor{BLUE}{rgb}{0,0,1}
 \definecolor{CYAN}{cmyk}{1,0,0,0}
 \definecolor{MAGENTA}{cmyk}{0,1,0,0}
 \definecolor{YELLOW}{cmyk}{0,0,1,0}
}

\usepackage{babel}

\newcommand{\even}{\text{even}}
\newcommand{\odd}{\text{odd}}

\newcommand{\norme}[1]{\left\Vert #1\right\Vert}

\makeatother

\usepackage{babel}
\begin{document}

\title{Thermalization in a 1D Rydberg gas:\\ validity of the microcanonical
ensemble hypothesis}

\author{Ruben Y. Cohen, Etienne Brion, Frédéric Grosshans }

\affiliation{Laboratoire Aimé Cotton, CNRS, Université Paris-Sud, ENS Cachan,
Université Paris-Saclay, 91405 Orsay Cedex, France.}
\begin{abstract}
We question the microcanonical hypothesis, often made to account for
the thermalization of complex closed quantum systems, on the specific
example of a chain of two-level atoms optically driven by a resonant
laser beam and strongly interacting via Rydberg-Rydberg dipole-dipole
interactions. Along its (necessarily unitary) evolution, this system
is indeed expected to thermalize, \emph{i.e.\@} observables, such as
the number of excitations, stop oscillating and reach equilibrium-like
expectation values. The latter are often calculated through assuming
the system can be effectively described by a thermal-like microcanonical
state. Here, we compare the distribution of excitations in the chain
calculated either according to the microcanonical assumption or through
direct exact numerical simulation. This allows us to show the limitations
of the thermal equilibrium hypothesis and precise its applicability
conditions. 
\end{abstract}


\maketitle

\section{Introduction}

\label{sec:Intro}

Due to their large dipole moments \cite{G94}, Rydberg atoms experience
strong long-range dipole-dipole interactions. Within the last decade,
this feature has been put forward as the key ingredient of different
promising atomic quantum-processing scenarios \cite{SWM10}. For instance,
Rydberg-Rydberg interactions can be used to perform two-qubit logic
operations in individual-atom systems by shifting a transition off-resonance
in an atom, depending on the internal state of another atom in its
immediate neighborhood \cite{JCZ00,PRS02,SW05}. In a mesoscopic ensemble,
dipole-dipole interactions are able to inhibit transitions into collective
states that contain more than one Rydberg excitation, thus leading
to the so-called Rydberg blockade. First predicted in \cite{LFC01},
this phenomenon was locally observed in laser-cooled atomic systems
\cite{TFS04,SRA04,CRB05,AVG98,VVZ06} and could in principle be used
in the future to manipulate and entangle collective excitation states
of mesoscopic ensembles of cold atoms which could therefore be run
as quantum processors \cite{LFC01,BMM07,BMS07,BPS08} or repeaters
\cite{ZMH10,HHH10,BCA12}. Rydberg atomic ensembles are also investigated
on quantum non-linear optical purposes \cite{WA13}: converting photons
into so-called Rydberg polaritons which strongly interact through
dipole-dipole interaction, it seems indeed possible to generate giant
non-linearities in the quantum regime, \emph{i.e.\@} to effectively
implement photon-photon interactions \cite{PFL12,PBS12,MSP13}.

The exact calculation of the time-dependent state of an ensemble of
atoms resonantly laser-driven towards a Rydberg level constitutes
a highly non-trivial coupled many-body problem. Such a complex system,
however, often shows an effective thermalization behavior 
\cite{LOG10,AL12,BMFALA13}: 
observables, such as the number of Rydberg excitations, indeed tend
to quasistationary values which can be computed assuming the system
is in a thermal equilibrium state, either in the canonical \cite{LOG10}
or microcanonical \cite{AL12,BMFALA13} ensembles. Considering the
same system as in \cite{AL12,BMFALA13}, we compare the predictions
of the microcanonical ensemble assumption to a numerical simulation
of the unitary evolution of a $100$-particle system, confirmed by
a simplified analytical treatment. The discrepancies we observe allow
us to show the limitations of the equilibrium hypothesis and to precise
its applicability conditions.

The paper is organized as follows. In Sec.~\ref{sec:PhysicalModel},
we present the physical system and simplified model. In Sec.~\ref{sec:Thermal},
we give an analytical description of the distribution of excitations
according to the microcanonical ensemble. In Sec.~\ref{Calculations},
we numerically compute the distribution of excitations and apply a
simplified analytical treatment which allows us to satisfactorily
reproduce the results of the full simulation, in the regime of at
most 2 excitations. In Sec.~\ref{sec:Comparison}, we discuss and
compare the results obtained according to the different approaches,
before concluding in Sec.~\ref{sec:Conclusion}.

\section{Model and approximations}

\label{sec:PhysicalModel}

We consider a system of $N$ identical atoms located along a line
of length $L$. The Hilbert space of each atom is assumed to be restricted
to the ground state $\left|g\right\rangle $ and a highly excited
(so-called) Rydberg state $\left|r\right\rangle $. In the ensemble
``vacuum state'' $\left|\varnothing\right\rangle $, all atoms are
in the ground state: $\left|\varnothing\right\rangle \equiv\left|g\dots g\right\rangle $.
Denoting by $\sigma_{+}\equiv\left|e\right\rangle \left\langle g\right|$
and $\sigma_{-}\equiv\sigma_{+}^{\dagger}$ the usual raising and
lowering operators for a two-level atom, one defines the ensemble
state $\left|i\right\rangle \equiv\sigma_{+}^{i}\left|\varnothing\right\rangle =\left|g_{1},\dots,g_{i-1},r_{i},g_{i+1},\dots,g_{N}\right\rangle $
in which the $i^{\text{th}}$ atom is Rydberg-excited while the others are
in the ground state. In the same way, one can define the doubly excited
state $\left|i,j\right\rangle \equiv\sigma_{+}^{\left(i\right)}\sigma_{+}^{\left(j\right)}\left|\varnothing\right\rangle $
which contains only two Rydberg excitations at positions $i$ and
$j$, and more generally any arbitrary multiply excited state $\left|i,j,k,\dots\right\rangle \equiv\sigma_{+}^{\left(i\right)}\sigma_{+}^{\left(j\right)}\sigma_{+}^{\left(k\right)}\dots\left|\varnothing\right\rangle $.

The atomic ensemble is subject to a laser beam which resonantly drives
the transition $\left|g\right\rangle \leftrightarrow\left|r\right\rangle $:
in the rotating wave approximation, this process is simply described
by the Hamiltonian $H_{L}=\hbar\Omega\sum_{k=1}^{n}\left(\sigma_{+}^{k}+\sigma_{-}^{k}\right)$,
where $\Omega$ denotes the laser Rabi frequency. Moreover, when lying
in their Rydberg state, two atoms interact through the (strong) dipole-dipole
interaction (this interaction is negligible when at least one atom
in the pair is in the ground state): the corresponding Hamiltonian
is 
\begin{equation}
V_{dd}=\hbar C_{6}\sum_{k\neq m}\frac{n_{m}n_{k}}{d(m,k)^{6}}\label{V_dd}
\end{equation}
where $C_{6}$ is the van der Waals interaction coefficient, $n_{k}\equiv\sigma_{+}^{k}\sigma_{-}^{k}$
the projector onto the Rydberg state for the $k^{\text{th}}$ atom and $d(m,k)$
is the distance between the $m^{\text{th}}$ and the $k^{\text{th}}$ atoms.
Finally,
the full Hamiltonian governing the dynamics of the system is 
\begin{align}
H & =H_{L}+V_{dd}\label{Ham}
\end{align}

Starting in the ensemble vacuum state $\left|\varnothing\right\rangle $,
in the absence of interatomic interactions, each atom in the sample
would independently undergo Rabi oscillations. Because of dipole-dipole
interactions, atoms actually get entangled during their evolution,
according to the so-called Rydberg blockade phenomenon \cite{LFC01}.
To understand this mechanism, let us first consider the simple case
of two atoms. If they are ``close enough'' so that their dipole-dipole
interaction overwhelms the laser Rabi frequency, their simultaneous
excitation into the Rydberg state becomes impossible since the doubly
excited state is strongly shifted out of resonance. As a rule of thumb,
one can define the typical distance $R_{b}$, called the blockade
radius, at which the blockade starts being effective as the distance
for which the van der Waals interaction becomes comparable with the
laser excitation, \emph{i.e} $R_{b}\approx\left(\frac{C_{6}}{\Omega}\right)^{\frac{1}{6}}$.
Now turning to the full sample, it is clear that dipole-dipole interactions
forbid the system to explore its full Hilbert space: too off-resonant
configurations, \emph{i.e.\@} ensemble states in which two Rydberg excited
atoms are closer than the radius $R_{b}$, will indeed never be substantially
populated. In other words, due to the Rydberg blockade the system
is bound to essentially evolve in the subspace of ``allowed states''
in which excited atoms are separated at least by $R_{b}$ (Note that
in a 3D geometric arrangement, each Rydberg excited atom creates an
``exclusion'' sphere of radius $R_{b}$ often called a ``Rydberg
bubble'').

Though simple in its form, the Hamiltonian Eq.~\eqref{Ham} leads
to complex many-body dynamics. In particular, besides the Rydberg
blockade phenomenon qualitatively described above, it was shown to
yield thermalization effects \cite{LOG10}. The full computation of
the dynamics is intractable for large numbers of atoms and one must
resort to approximations. Following \cite{LOG10}, we shall make the
hardcore Rydberg sphere assumption, that is we shall merely discard
all atomic configurations in which two Rydberg excitations are closer
than $R_{b}$, while keeping the others; moreover, we shall make the
simplistic approximation that in the allowed subspace the dipole-dipole
Hamiltonian is zero. In this approximation the full Hamiltonian therefore
becomes 
\begin{equation}
H\approx\hbar\Omega\sum_{k=1}^{n}(\tilde{\sigma}_{+}^{k}+\tilde{\sigma}_{-}^{k})\label{Happrox}
\end{equation}
where $\tilde{\sigma}_{+}^{k}$ is the raising operator of the $k^{\text{th}}$
atom restricted to the allowed configuration subspace, \emph{i.e.\@}
the operator which excites the $k^{\text{th}}$ atom into the Rydberg
state provided that no other Rydberg atom is in the range $R_{b}$.

\section{Thermalized state: Analytical results from the microcanonical ensemble
assumption }

\label{sec:Thermal}
\
Analytical \cite{AGL12,JAG13} and numerical \cite{LOG10} investigations
of the approximate Hamiltonian Eq.~\eqref{Happrox} both predict
thermalization to occur in the system. Intuitively, this phenomenon
results from the destructive interferences between different frequency
components of the evolved vector state: for large times, due to the
complexity of the Hilbert space and the high connectivity of the basis
states, observables, such as the number of Rydberg-excited atoms,
are expected to stop oscillating and tend to quasistationary values.
According to the microcanonical ensemble assumption \cite{AL12,BMFALA13},
these values can be accounted for by assuming an effective thermal
equilibrium-like state for the system which consists of an equiprobable
statistical mixture of all allowed states.

The common probability of all the components in this mixture is therefore
simply given by the inverse of the total number $\mathcal{N}$ of
allowed states. This number can be determined by summing all numbers
$\mathcal{N}(\nu)$ of allowed configurations with exactly
$\nu$ excitations, which, as we show below, can be calculated through
a straightforward combinatorial argument. From $\mathcal{N}(\nu)$,
one easily computes the average number of excitations $\left\langle \nu\right\rangle $
and, in the limit of a continuous distribution, one can even deduce
a simple expression of the spatial density of Rydberg excitations.

In this section, we present our analytical calculations in detail
and compare our results to numerical Monte Carlo simulations presented
in \cite{BMFALA13}.

\subsection{Number of allowed states and average excitation number: combinatorial
analysis}

The goal of this subsection is to compute the average number of Rydberg
excitations observed in the thermalized state according to the microcanonical
ensemble assumption. For sake of simplicity, we assume that the atoms
are located at the nodes of a regular 1D lattice of step $a$. The
distance between the $i^{\text{th}}$ and $j^{\text{th}}$ atoms is therefore $d(i,j)=a\left|i-j\right|$
while the total length of the line is given by $L=\left(N-1\right)a$.
The quantity $n_{b}\equiv\left\lfloor \frac{R_{b}}{a}\right\rfloor $,
where $\left\lfloor \cdot\right\rfloor $ denotes the lower integer
part, represents the minimal number of ground-state atoms which must
lie between two Rydberg excitations in an allowed atomic configuration
according to the hardcore Rydberg sphere assumption. Finally, we introduce
the real parameter $\Lambda\equiv\frac{L}{R_{b}}$. Adding one to its integer part
gives the maximum number of Rydberg excitations the sample can accommodate for: 
$\nu_{\max}=\left\lfloor \Lambda\right\rfloor + 1$.

\begin{figure}
\includegraphics[width=8cm]{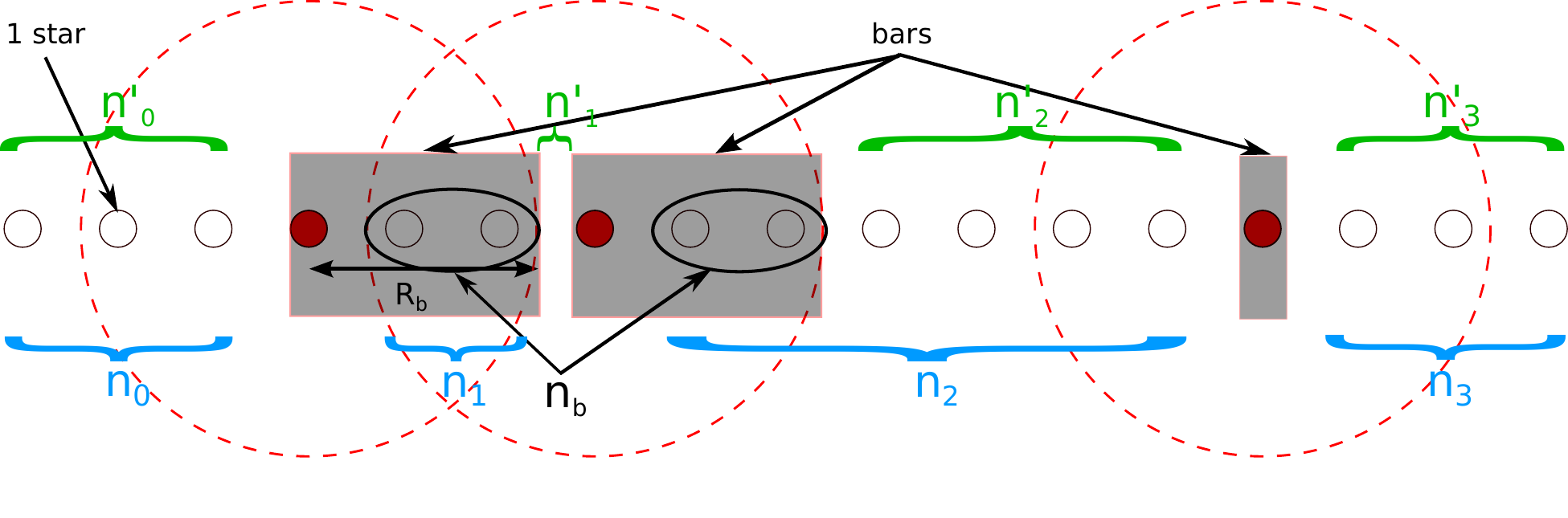}%

{\Large{}{}$\overbrace{\bigstar\bigstar\bigstar}^{n'_{0}}|\overbrace{\phantom{|}}^{n'_{1}}|\overbrace{\bigstar\bigstar\bigstar\bigstar}^{n'_{2}}|\overbrace{\bigstar\bigstar\bigstar}^{n'_{3}}$}
\protect\protect\caption{\label{fig:StarsBars}: Description of a configuration of the excitations
and mapping to the ``Stars and bars'' problem. An excited atom with
all the atoms on its right closer than $R_b$ correspond to a bar, except
for the last bar constituted only by the last excited atom. Each remaining
atom represent a star. The depicted configuration corresponds to $L=16$,
$N=17$, $2a\le R_{b}<3a$, $n_{b}=2$, $\{n_{k}\}_{k}=\{3,2,6,3\}$,
$\{n'_{k}\}_{k}=\{3,0,4,3\}$.}
\end{figure}

To begin with, we compute the number of allowed states which comprise
a given number of excitations $\nu$. In such a state, the $\nu$
Rydberg excitations split the sample into $\left(\nu+1\right)$ groups
of $n_{k=0,\dots,\nu}$ ground-state atoms (see Fig.~\ref{fig:StarsBars}),
with the convention that the zeroth and $\nu^{\text{th}}$ groups are on the
left and the right of the leftmost and rightmost excited atoms, respectively,
and allowing $n_{0}$ and $n_{\nu}$ to be zero. The state indeed
corresponds to an allowed configuration if it satisfies the hardcore
Rydberg sphere condition, \emph{i.e.\@} $n_{k}\geq n_{b}$ for $1\leq k\leq\left(\nu-1\right)$,
under the prescription $\sum_{k=0}^{\nu}n_{k}=N-\nu$: finding the
number of allowed states with $\nu$ excitations is therefore equivalent
to computing the number of sets of integers $\left\{ n_{k=0,\dots,\nu}\right\} $
which satisfy the two previous conditions. A slight modification in
the formulation of this problem turns it into a standard combinatorial
calculation as we shall now show. We first note that an allowed atomic
configuration can be uniquely determined by the alternative set of
numbers $\left\{ n{}_{k}'\right\} $ defined by 
\begin{align*}
n_{0}' & \equiv n_{0}\\
n_{k}' & \equiv n_{k}-n_{b}\;\text{for }1\leq k\leq\nu-1\\
n_{\nu}' & \equiv n_{\nu}
\end{align*}
which satisfy the conditions $n_{k}'\geq0$ and $\sum_{k=0}^{\nu}n'_{k}=N-1-(\nu-1)(n_{b}+1)$.
This change of variables suggests to associate the original atomic
arrangement with an abstract linear distribution of $\left[N-1-(\nu-1)(n_{b}+1)\right]$
``stars'' split by $\nu$ ``bars'' into $\left(\nu+1\right)$
groups labelled by $k=0,\dots,\nu$ and respectively comprising $n_{k}'$
elements. As shown in Fig.~\ref{fig:StarsBars}, the first $\left(\nu-1\right)$
bars symbolize the first $\left(\nu-1\right)$ Rydberg excited atoms
with their first $n_{b}$ (ground-state) right neighbors, while the
last bar represents the last Rydberg excited atom only; stars then
simply stand for the remaining ground state atoms. Calculating the
number $\mathcal{N}(\nu)$ of such configurations is a
standard combinatorial problem whose solution is given by the binomial
coefficient 
\begin{align}
\mathcal{N}(\nu) & =\binom{N-(\nu-1)n_{b}}{\nu}\nonumber \\
 & =\frac{N^{\nu}}{\nu!}\prod_{i=0}^{\nu-1}\left(1-\frac{(\nu-1)n_{b}+i}{N}\right)\label{eq:Nnudiscrete}
\end{align}
Note that $\mathcal{N}(\nu)=0$ when $\nu-1\geq\frac{N}{\left\lfloor R_{b}\right\rfloor +1}=\frac{L+1}{\left\lceil R_{b}\right\rceil }$.
In the limit of large $N$ and $R_{b}$, this essentially means that
we only have to consider configurations with a number of excitations
smaller than $\nu\lesssim\Lambda$. In this limit, when $\Lambda\ll R_b,N$,
we can approximate equation \eqref{eq:Nnudiscrete} by 
\begin{equation}
\mathcal{N}(\nu)=\frac{N^{\nu}}{\nu!}\left[1-\frac{\nu-1}{\Lambda}\right]_{+}^{\nu}+O(N^{\nu-1}),\label{eq:apNnudiscrete}
\end{equation}
where $[x]_{+}^{\nu}=0$ if $x\le0$ and $[x]_{+}^{\nu}=x^{\nu}$
if $x\ge0$.

\begin{figure}
\centering{}\includegraphics[width=8cm]{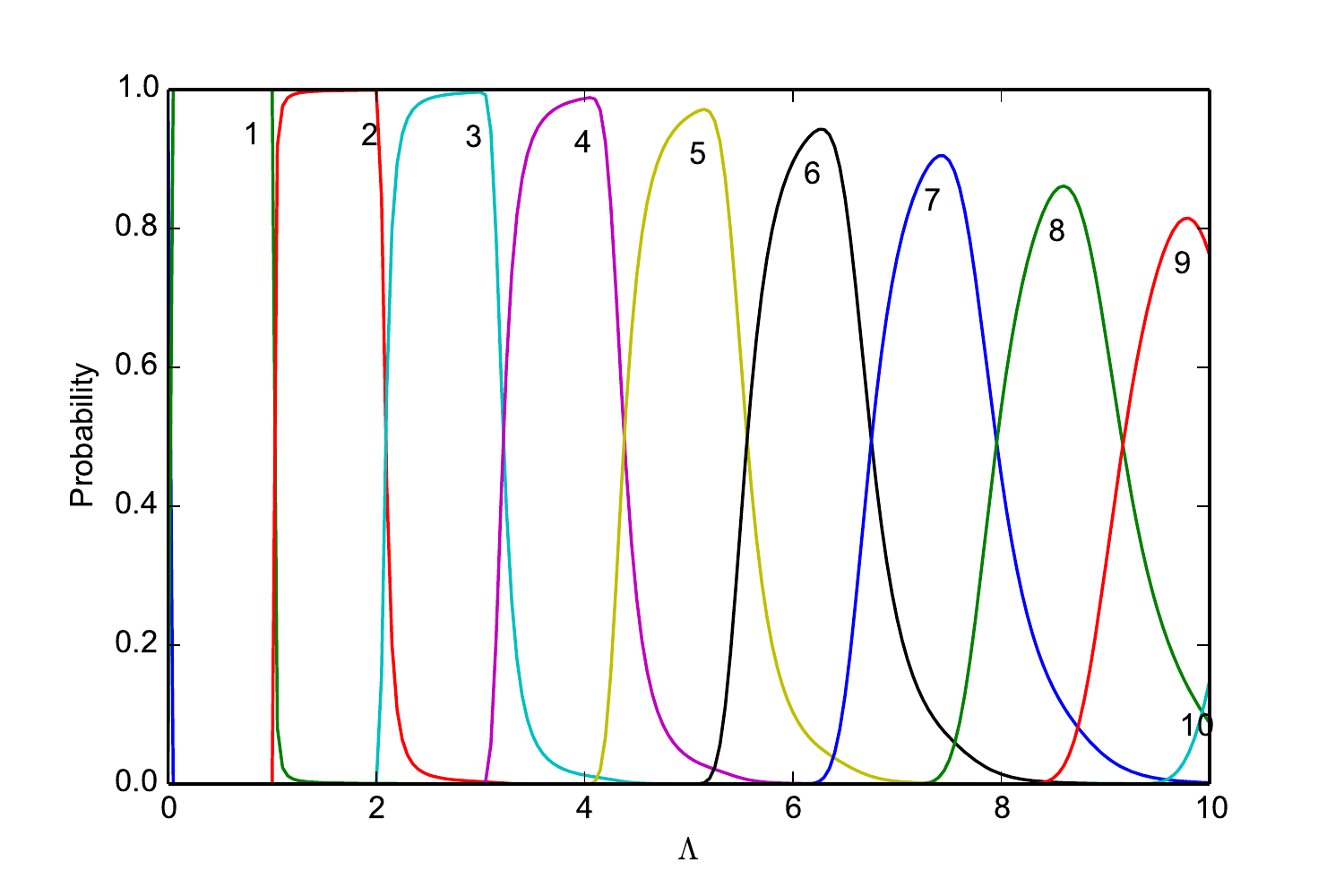}\protect\caption{\label{fig:ExcProba} Probability to have $\nu$ excitation considering
the microcanonical ensemble as a function of $\Lambda$ for $N=10^{4}$.
The successive peaks correspond to increasing value of $\nu$. For
example, $P(\nu=2)$ is close to 1 when $\Lambda$ is between 1 and
2. }
\end{figure}

\begin{figure}
\centering{}\includegraphics[width=8cm]{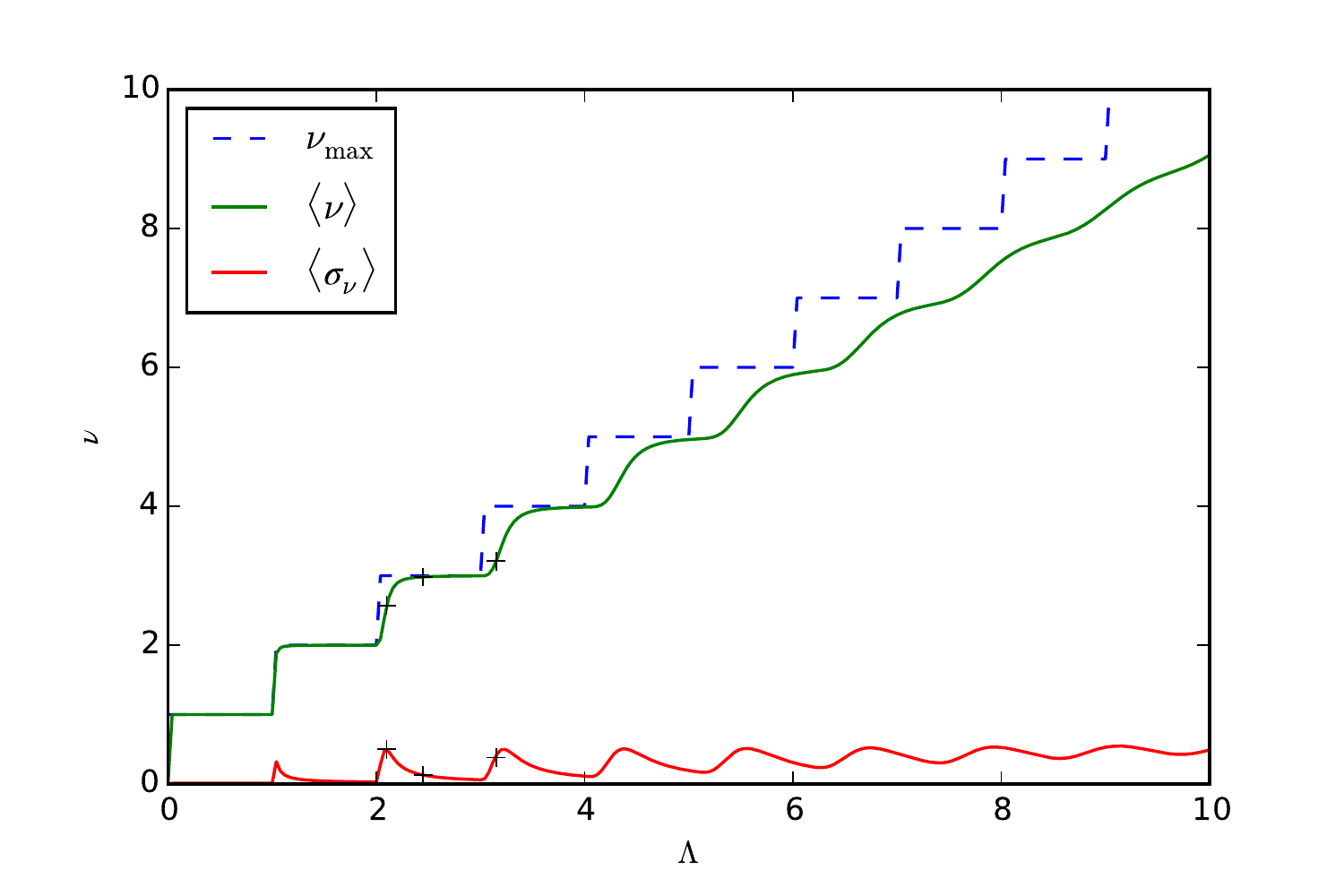}\protect\caption{\label{fig:AvrgNu} Average $\nu$ and standard deviation in the microcanonical
predictions as a function of $\Lambda$ for $N=10^{4}$. The crosses
correspond to the Monte-Carlo results from \cite[fig 2c]{BMFALA13}.}
\end{figure}

From $\mathcal{N}(\nu)$, one easily computes the total number of
allowed configurations $\mathcal{N}=\sum_{\nu}\mathcal{N}(\nu)$,
the probability to have $\nu$ excitations in the sample $\mathcal{P}(\nu)=\mathcal{N}(\nu)/\mathcal{N}$
and the average excitation number $\langle\nu\rangle=\sum_{\nu}\nu\mathcal{P}(\nu)$
as well as its standard deviation $\left\langle \sigma_{\nu}\right\rangle $
as a function of $\Lambda$. The family of curves $\left\{ \mathcal{P}(\nu),\nu=0,1,\dots\right\} $
is plotted on Fig.~\ref{fig:ExcProba} as a function of $\Lambda$
for $N=10^{4}$ and on Fig.~\ref{fig:Probabilities} for $N=10^{2}$;
$\left\langle \nu\right\rangle $ and $\left\langle \sigma_{\nu}\right\rangle $
are represented on Fig.~\ref{fig:AvrgNu} as functions of $\Lambda$
for $N=10^{4}$. Our results show perfect quantitative agreement with
\cite{BMFALA13}, as detailed in Appendix~\ref{App:Comparison}

\subsection{Spatial density of Rydberg excitations}

We can go further in our analysis and compute how Rydberg excitations
are distributed along the line in average. Calculations turn to be
much easier in the limit of a homogeneous and continuous atomic distribution,
of constant linear density $\delta\equiv\frac{1}{a}$ which is a good
approximation of our model when $R_{b},L\gg a$.

Let us denote by $\mathcal{N}(\nu,l)$ the number of configurations
with $\nu$ excitations on a line of length $l$ with the density
$\delta$. We have 
\begin{align*}
\forall l\ge0,\;\mathcal{N}(0,l) & =1\\
\forall l<0,\forall\nu,\;\mathcal{N}(\nu,l) & =0
\end{align*}
With these notations, $\mathcal{N}(\nu)=\mathcal{N}(\nu,L)$
and if $\nu>0$, the number of configurations with the leftmost excited
atom at position $x$ is given by $\mathcal{N}(\nu-1,L-R_{b}-x)$.
Integrating over $x$, we get the recurrence relation 
\begin{equation}
\mathcal{N}(\nu+1,L)=\int_{0}^{L}dx\:\delta\:\mathcal{N}(\nu,L-R_{b}-x),
\end{equation}
and 
\begin{align}
\mathcal{N}(\nu,L) & =\frac{\delta^{\nu}}{\nu!}\left[l-(\nu-1)R_b\right]_{+}^{\nu}\nonumber \\
 & =\frac{N^{\nu}}{\nu!}\left[1-\frac{\nu-1}{\Lambda}\right]_{+}^{\nu}\label{eq:Nnu2}
\end{align}
which is consistent with Eq.~\eqref{eq:apNnudiscrete}.

\begin{figure}
\centering{}\includegraphics[width=8cm]{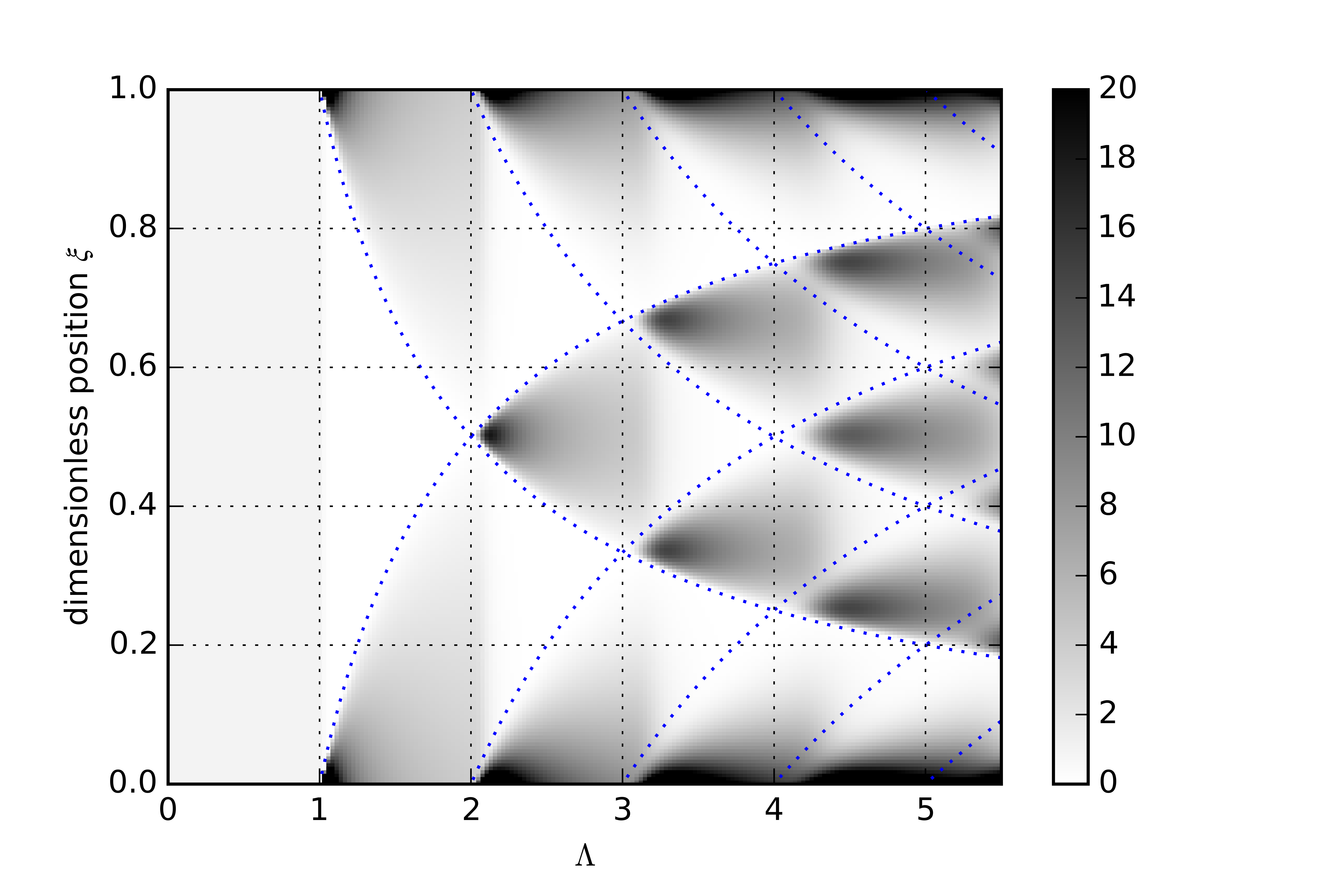}\protect\caption{\label{fig:ExcDistribution} Probability distribution of Rydberg excitations
along the chain as a function of $\Lambda$ and the position for $N=10^{4}$
atoms. This figure quantitatively reproduces the Monte-Carlo simulation 
of \cite[fig 2a]{BMFALA13}, as detailed in Appendix~\ref{App:Comparison}.}
\end{figure}

The probability density to have the $n^{\text{th}}$ excited atom out of $\nu$
at the position $x$ is: 
\begin{align}
  p(\nu,n,x)= 
    & \delta\:\frac{\mathcal{N}(n-1,x-R_b)\times\mathcal{N}(\nu-n,L-R_b-x)}{\mathcal{N}(\nu)} 
    \notag\\
  = & \frac{\nu!\left[\xi-\tfrac{n-1}{\Lambda}\right]_{+}^{n-1}%
          \left[1-\xi-\tfrac{\nu-n}{\Lambda}\right]_{+}^{\nu-n}}%
    {\left(n-1\right)!\left(\nu-n\right)!\left[1-\tfrac{\nu-1}{\Lambda}\right]_{+}^{\nu}}
    \label{eq:excdensity}
\end{align}
where we introduced the normalized dimensionless position $\xi\equiv\frac{x}{L}$.
Note that it does not depend on $N$; as seen above, however, $N$
plays a role in the global probability for having $\nu$ excitations.
This allows to plot the spatial distribution of excitation $\mathcal{P}\left(x\right)=\sum_{\nu}\sum_{n\leq\nu}p\left(\nu,n,x\right)$,
as in Fig.~\ref{fig:ExcDistribution} for $N=10^{4}$, which quantitatively
agrees with the Monte-Carlo simulations provided in \cite[fig 2a]{BMFALA13,BMFALA13data}
(see Appendix~\ref{App:Comparison}).
The spatial distribution of excitations is also plotted in Fig.~\ref{fig:exc_proba_1,5_}
(red curve) for $N=100$ and $\Lambda=1.5$.

\section{Numerical and simplified analytical calculations of the thermalized
state \label{Calculations}}

\begin{figure}
\centering{}\includegraphics[width=8cm]{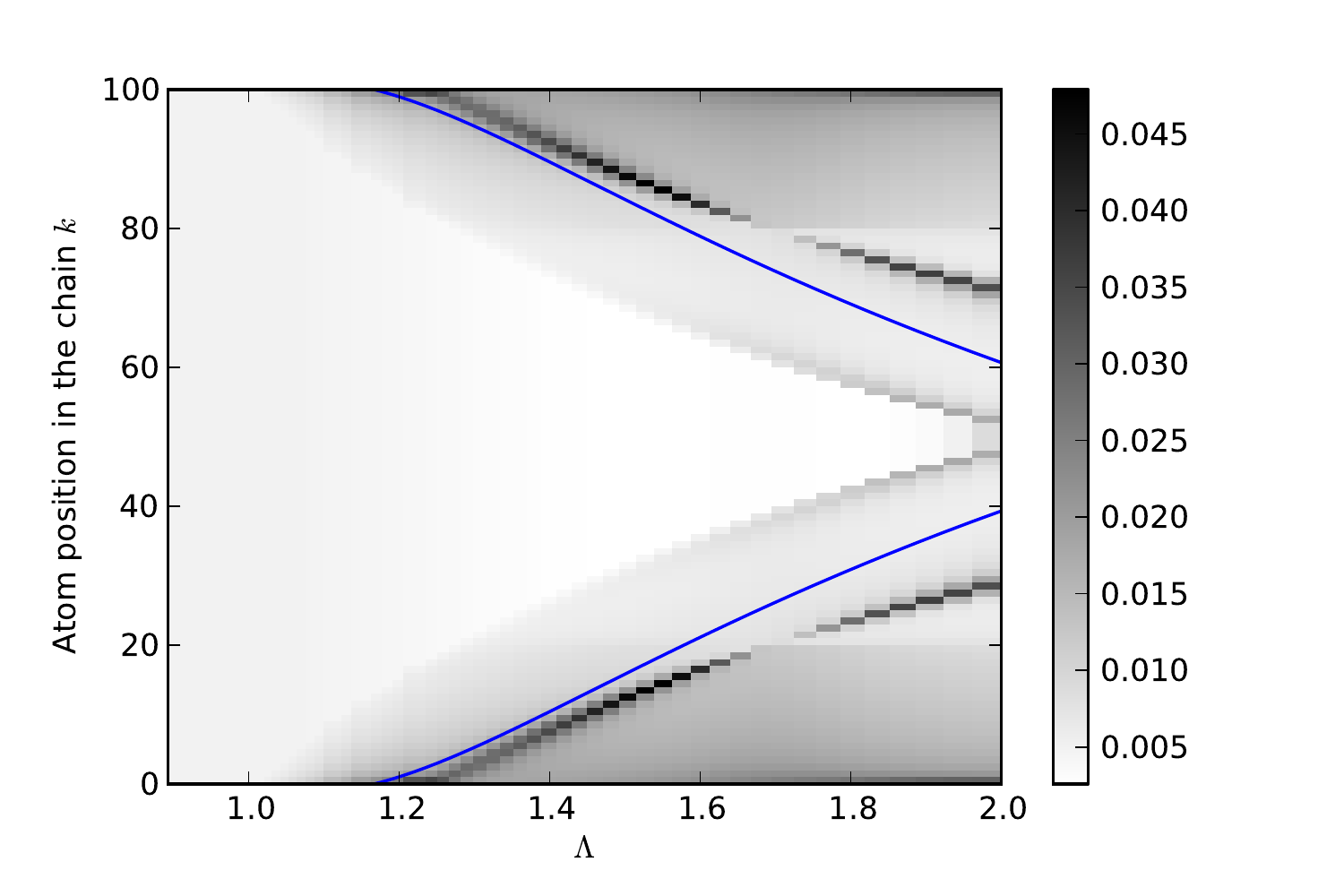}\protect\caption{\label{fig:localisationExc2}: Numerically computed probability distribution
$\mathcal{P}_{k}$ of Rydberg excitations along the chain, as a function
of $\Lambda$ and the position (see section \ref{sec:Numerics}) for
$N=100$ atoms. The blue curve is the predicted position of the excitation
peak by our simplified analytical treatment.}
\end{figure}

In this section, we present the results we obtained through a direct
numerical calculation of the thermalized state averages and recover
some of their interesting features through approximately diagonalizing
the Hamiltonian in a conveniently truncated basis.

\subsection{Numerical calculation of the thermalized state \label{sec:Numerics}}

Our numerical method consists in time-averaging the observables of
interest: if the average is performed on a very long (ideally infinite)
time, the obtained average must indeed coincide with the thermalized
value. Due to computational complexity, we restricted our study to
a modest system of $N=100$ atoms equally spaced along the chain.
Even with this relatively small value of $N$, the dimension $2^{N}$
of the complete Hilbert space makes the full dynamical treatment intractable.
We therefore restricted ourselves to the regime $\Lambda<2$, \emph{i.e.\@}
the chain is shorter than two Rydberg radii ($L\leq2R_{b}$) and the
maximum number of excitations distributed along the chain is $2$.
We only need to take into account the states allowed by the Rydberg
blockade whose number is given by Eq.~\eqref{eq:Nnu2}:
$\sum_{\nu\leq2}\mathcal{N}(\nu)\simeq\mathcal{N}(\nu=2)=\frac{N^{2}}{2}\left[1-\frac{1}{\Lambda}\right]_{+}^{2}$.
We generate this set of allowed states through an arborescent
search starting from $\left|\varnothing\right\rangle $ and adding
allowed excitations.

\begin{figure}
\centering{}\includegraphics[width=8cm]{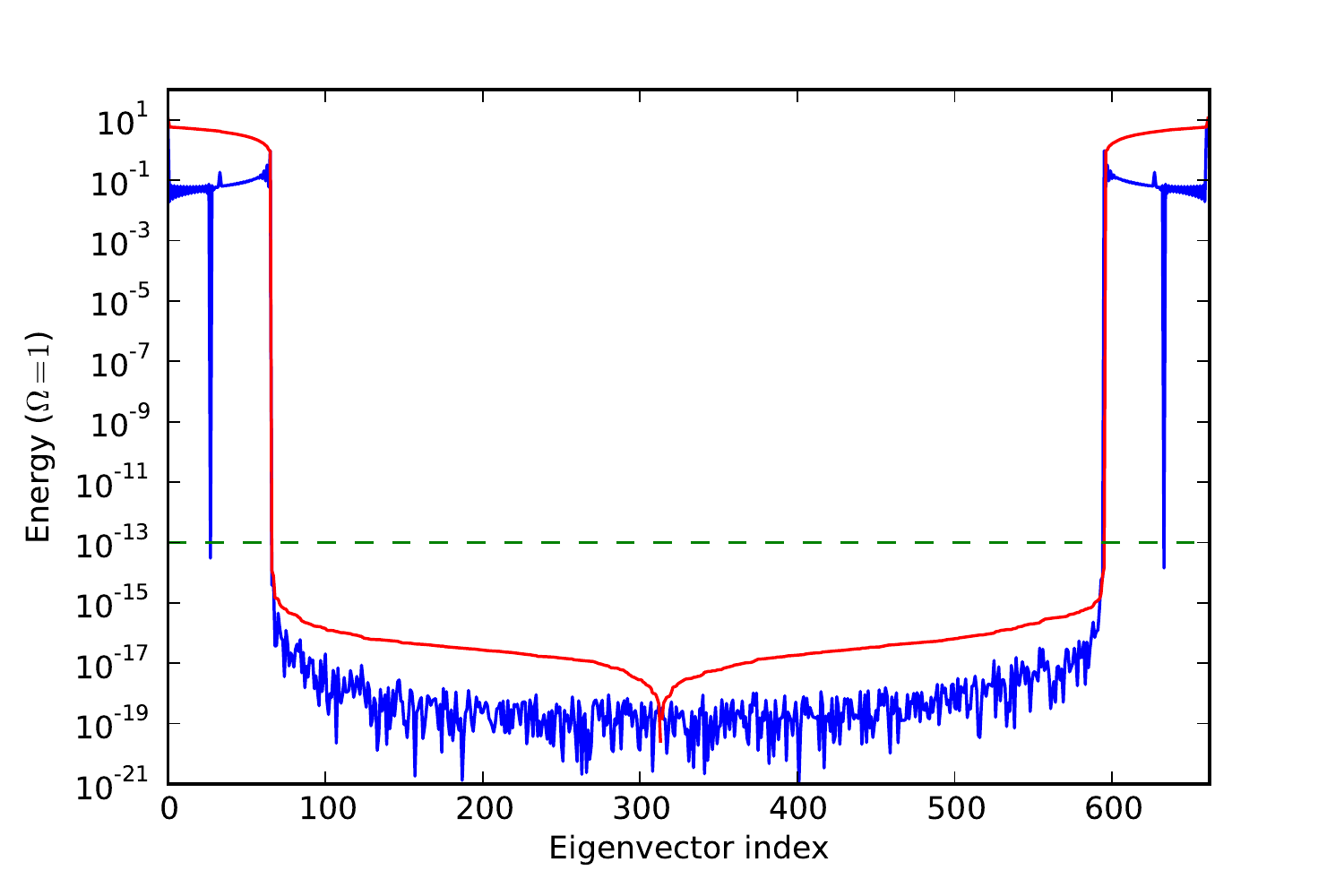}\protect\caption{\label{fig:DiffEnergy} In red: $|E_{n}|$ versus $\left(n,\alpha_{n}\right)$ with $\Lambda = 1.5$
(eigenstates are arranged in increasing order of their eigenenergy).
In blue: difference between two successive eigenvalues, \emph{i.e.\@}
$\left|E_{n+1}-E_{n}\right|$. The dashed line shows the degeneracy
limit: below this line, any values can be assumed to be zero, up to numerical artifacts (see text). 
}
\end{figure}

In this subspace, we numerically diagonalize the Hamiltonian of
Eq.~\eqref{Happrox}, yielding the (possibly degenerate) eigenenergies
$E_{n}$ and the associated eigenvectors 
$\ket{\psi_{n}^{\left(\alpha_{n}\right)}} $
where $\alpha_{n}=1\dots d_{n}$, $d_{n}$ are the degeneracy index
of the eigenenergy $E_{n}$. Fig.~\ref{fig:DiffEnergy} presents the
numerical results of the diagonalization of $H$: more explicitly,
the red curve shows the absolute value $\left|E_{n}\right|$ versus
the rank of the corresponding eigenvectors 
$\ket{\psi_{n}^{\left(\alpha_{n}\right)}}$,
arranged in increasing order of their eigenenergy; the blue curve
represents the energy difference between two successive eigenvectors
and therefore allows to check degeneracy. We take as a numerical criterion
that two energies coincide when their difference is less than $10^{-13}\Omega$,
consistent with the precision of IEEE 754 floating-point arithmetics.
One first observes a wide central area corresponding to the highly
degenerate eigenenergy $E\approx0$; in addition, on both
sides of the spectrum, there exist two pairs of eigenstates with degenerate
energies.

If the system is initially prepared in 
$\ket{\Psi\left(0\right)} \equiv 
 \sum_{n,\alpha_{n}}c_{n}^{\alpha_{n}}\ket{\psi_{n}^{\left(\alpha_{n}\right)}}$,
its state at time $t$ is given by 
$\ket{\Psi\left(t\right)} =
  \sum_{n,\alpha_{n}}c_{n}^{\alpha_{n}}e^{-\mathrm{i}\frac{E_{n}}{\hbar}t}
  \ket{\psi_{n}^{\left(\alpha_{n}\right)}} $.
The time-averaged probability $\mathcal{P}_{k}$ to have a Rydberg
excitation in site $k$ is therefore given by $\mathcal{P}_{k}=\mathrm{Tr}\left[\bar{\rho}\sigma_{rr}^{\left(k\right)}\right]$
where the average state $\bar{\rho}$ is 
\begin{align}
\bar{\rho} & =\overline{\left|\Psi\left(t\right)\right\rangle \left\langle \Psi\left(t\right)\right|}\nonumber \\
 & =\sum_{m,\alpha_{m}}\sum_{n,\beta_{n}}c_{m}^{\alpha_{m}}\left(c_{n}^{\beta_{n}}\right)^{*}\left|\psi_{m}^{\left(\alpha_{m}\right)}\right\rangle \left\langle \psi_{n}^{\left(\beta_{n}\right)}\right|\times\overline{e^{-\mathrm{i}\frac{E_{m}-E_{n}}{\hbar}t}}\nonumber \\
 & =\sum_{n,\alpha_{n},\beta_{n}}c_{n}^{\alpha_{n}}\left(c_{n}^{\beta_{n}}\right)^{*}\left|\psi_{n}^{\left(\alpha_{n}\right)}\right\rangle \left\langle \psi_{n}^{\left(\beta_{n}\right)}\right|\label{EqState}.
\end{align}
We have used the time average $\overline{e^{-\mathrm{i}\frac{E_{m}-E_{n}}{\hbar}t}}=\delta_{mn}$
to simplify the double sum. 

The probability distribution $\mathcal{P}_{k}$ is represented on
Fig.~\ref{fig:localisationExc2} as a function of $\Lambda$. For
$\Lambda\gtrsim1.2$, two one-atom-wide black lines appear, revealing
a strong localization of Rydberg excitations. In the next subsection,
we account for this phenomenon through the approximate diagonalization
of the Hamiltonian in a conveniently truncated basis.

\subsection{Simplified analytical treatment \label{SimpleAnalyticalTreatment}}

We are now looking for a simple description of the system which would
retain the basic physical features of the model, in particular the
localization effect observed in the previous subsection. To this end,
we shall try and restrict the basis of the whole Hilbert space to
only the relevant states, \emph{i.e.\@} those which get significantly
populated during the evolution.

\begin{figure}
\centering{}\includegraphics[width=8cm]{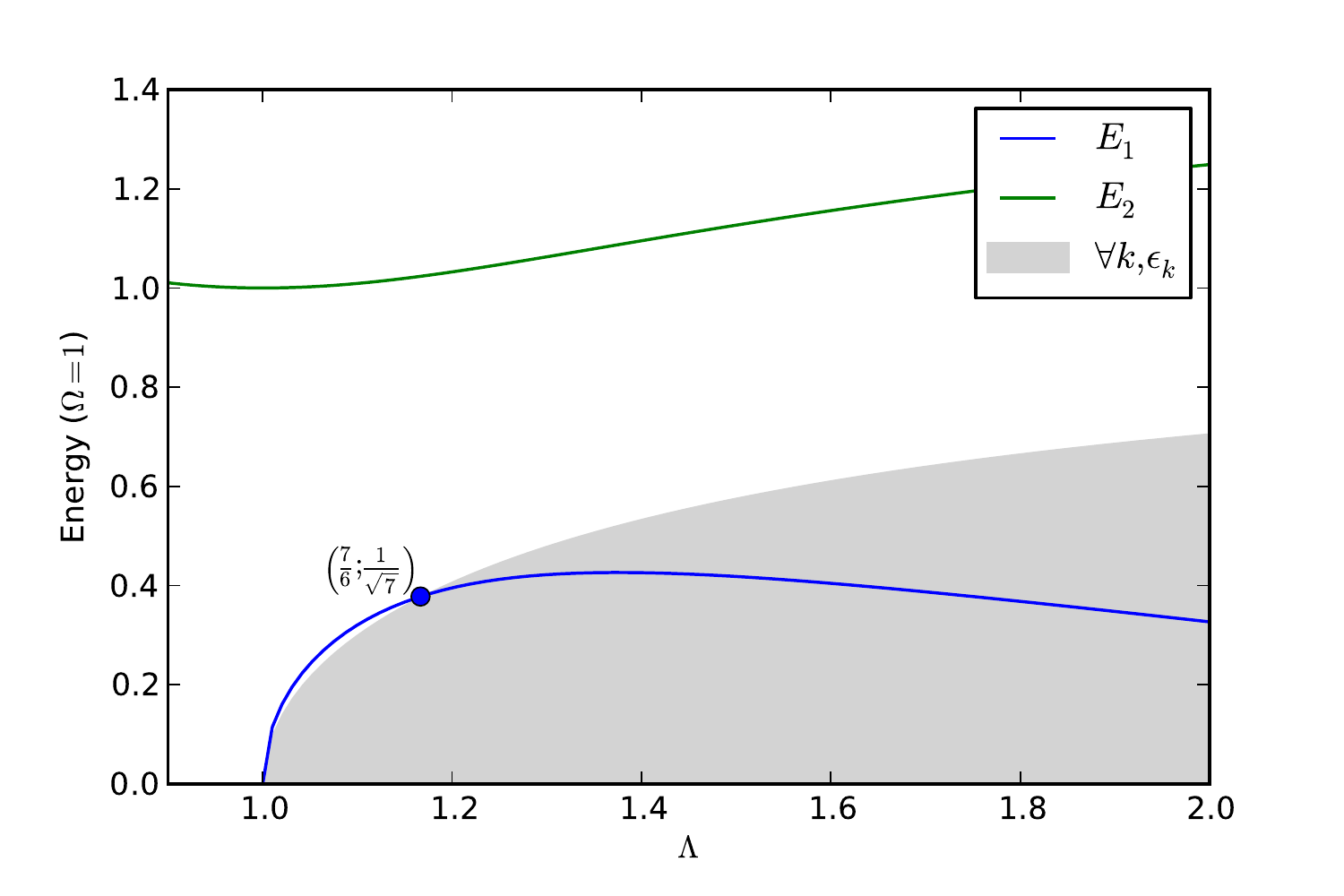}\protect\caption{\label{fig:EvsLambda} $E_{1},E_{2}$ and $\varepsilon_{k}$ as functions
of $\Lambda$, computed by our simplified analytical treatment for
$N=100$. The values of $\varepsilon_{k}$ form a quasi-continuum.
As discussed in the text, localization peaks arise when a resonance
takes place,\emph{i.e.\@} when there exists a value $k=K$ such that
$\varepsilon_{K}=E_{1}$. This happens for $\Lambda\geq\frac{7}{6}$
as can be shown analytically (see Appendix~\ref{App:Localization}) and graphically checked
on the present Figure. }
\end{figure}

As a first try, we consider the four-dimensional basis $\left\{ \left|\varnothing\right\rangle ,\left|\phi_{1}\right\rangle ,\left|\phi_{2}\right\rangle ,\left|\phi_{3}\right\rangle \right\} $
defined by
\begin{align}
  \left|\phi_{1}\right\rangle  
  & \equiv\frac{H\left|\varnothing\right\rangle }%
		{\left\Vert H\left|\varnothing\right\rangle \right\Vert } \label{state1}\\
	\left|\phi_{2}\right\rangle  
	& \equiv\frac{\Pi_{2}H\left|\phi_1\right>}%
			{\left\Vert\Pi_{2}H\left|\phi_1\right>\right\Vert}
		=\frac{\Pi_{2}H^{2}\left|\varnothing\right\rangle }%	
			{\left\Vert \Pi_{2}H^{2}\left|	\varnothing\right\rangle \right\Vert }%
	 \label{state2}\\
	\left|\phi_{3}\right\rangle  &
	 \equiv\frac{H\left|\phi_2\right>}%
			{\left\Vert H\left|\phi_2\right>\right\Vert}
		=\frac{H\Pi_{2}H^{2}\left|\varnothing\right\rangle }%
			{\left\Vert H\Pi_{2}H^{2}\left|\varnothing\right\rangle \right\Vert  }%
	  \label{state3}
\end{align}
where $\Pi_{2}$ denotes the projector onto the subspace of states
with exactly two Rydberg excitations. The diagonalization of $H$
in this subspace yields four eigenstates $\left|\psi_{i=1,2}^{s=\pm}\right\rangle $
and eigenenergies $\pm E_{i=1,2}$, such that $H\left|\psi_{i}^{s}\right\rangle =s\times E_{i}\left|\psi_{i}^{s}\right\rangle $,
whose explicit expressions can be found in Appendix~\ref{App:DiagH2}. We conventionally
choose $E_{2}\geq E_{1}\geq0$. The eigenenergies $E_{i}$ are plotted
as functions of $\Lambda$ on Fig.~\ref{fig:EvsLambda}. Note that
for $\Lambda>1$, all four eigenenergies $\pm E_{i=1,2}$ take different
values, there is hence no degeneracy. 

Since the eigenstates $\left|\psi_{i}^{s}\right\rangle $ describe
configurations where excitations are delocalized (see Appendix~\ref{App:Localization}),
the probability $\mathcal{P}_{k}$ computed from the time-averaged
state Eq.~(\ref{EqState})
\begin{align*}
\bar{\rho} & =\sum_{i=1,2}\sum_{s=\pm}\left|c_{i}^{s}\right|^{2}\left|\psi_{i}^{s}\right\rangle \left\langle \psi_{i}^{s}\right|
\end{align*}
will not exhibit the observed strong localization effect. 
Note that the four eigenstates
$\left|\psi_{i=1,2}^{s=\pm}\right\rangle $ contribute to the statistical
mixture $\bar{\rho}$ with the respective weights 
$\left|c_{i}^{s}\right|^{2}\equiv\left|\Braket{\psi_{i}^{s}|\Psi(0)}\right|^{2}$
determined by the initial state vector 
$\ket{\Psi(0)}=\ket{\varnothing}$. 

To correctly account for the observed localization phenomenon, we
must therefore slightly extend the basis. To this end, we consider
the family of states $\left\{ \ket{\varphi_{k=1,\dots,N-n_b-1}^{s=\pm}}\right\} $
defined by
\begin{equation}
  \Ket{\varphi_{k=1,\dots,N-R_b-1}^{\pm}} \equiv
    \frac{\ket{\Phi_{k}^{\left(1\right)}} \pm \ket{\Phi_{k}^{\left(2\right)}}}%
         {\sqrt{2}} \label{state}
\end{equation}
with
\begin{align*}
\left|\Phi_{k}^{\left(1\right)}\right\rangle  & \equiv\frac{\left|k\right\rangle +\left|N-k\right\rangle }{\sqrt{2}}\\
\left|\Phi_{k}^{\left(2\right)}\right\rangle  & \equiv\sum_{l=0}^{N-n_b-k-1}\frac{\left|k,N-l\right\rangle +\left|N-k,l\right\rangle }{\sqrt{2\left(N-n_b-k\right)}}
\end{align*}
Note that $\ket{\Phi_{k}^{\left(1\right)}}$ describes
a configuration with exactly one Rydberg excited atom, localized either
at position $k$ or $\left(N-k\right)$; 
$\ket{\Phi_{k}^{\left(2\right)}} $
describes a configuration with two Rydberg excitations, one being
localized in $k$ or $\left(N-k\right)$ while the other is fully
delocalized along the chain. The states $\left|\varphi_{k}^{s}\right\rangle $
are therefore coherent superpositions of states with either one or
two excitations, one being localized with certainty either at position
$k$ or $\left(N-k\right)$.

The states $\left|\varphi_{k}^{s}\right\rangle $ are found to be
approximately orthogonal to $\left|\psi_{k}^{s}\right\rangle $, \emph{i.e.\@}
\[
\left\langle \varphi_{k}^{s}\middle|\psi_{k'}^{s'}\right\rangle =O\left(\frac{1}{\sqrt{N}}\right).
\]
They are, moreover, only very weakly coupled to $\left|\psi_{k}^{s}\right\rangle $
by the Hamiltonian, \emph{i.e.\@}
\begin{equation}
\left\langle \varphi_{k}^{s}\middle| H\middle|\psi_{k'}^{s'}\right\rangle =O\left(\frac{1}{\sqrt{N}}\right).\label{HamDiag}
\end{equation}
Finally, for any $k=1,\dots,\left(N-R_b-1\right)$ and $s=\pm$, one
has 
\[
\left\langle \varphi_{k}^{s}\left|H\right|\varphi_{k}^{s}\right\rangle =s\times\varepsilon_{k}.
\]
The expression of $\varepsilon_{k}$ is given in Appendix~\ref{App:Localization}.
Fig.~\ref{fig:EvsLambda} shows the quasi-continuum formed by the
different $\varepsilon_{k}$'s plotted as functions of $\Lambda$.

If the system starts in a superposition of $\ket{\Psi_{k}^{s}}$,
\emph{i.e.\@} $\ket{\psi\left(0\right)} =\sum_{k,s}c_{k}^{s}\ket{\psi_{k}^{s}}$,
one could be tempted, due to Eq.~(\ref{HamDiag}), to assume that
none of the states $\ket{\varphi_{k}^{s}}$ ever gets
substantially populated and to discard the whole family 
$\left\{\ket{\varphi_{k}^{s}} \right\} $
from our description. This would actually be incorrect: it may indeed
happen that, for a given $k=K$, $\left|\varphi_{K}^{s}\right\rangle $
becomes resonant with $\left|\psi_{1}^{s}\right\rangle $, \emph{i.e.\@}
$\varepsilon_{K}=E_{1}$ (as can be checked on Fig.~\ref{fig:EvsLambda},
such a resonance exists only for $\Lambda\geq\frac{7}{6}$; 
in Appendix~\ref{App:Localization}, this
result is also analytically deduced from the expressions of $E_{1}$
and $\varepsilon_{k}$). In that case, though very weak,
the coupling term $\left\langle \varphi_{K}^{s}\left|H\right|\psi_{1}^{s}\right\rangle $
strongly mixes the states $\left|\varphi_{K}^{s}\right\rangle $ and
$\left|\psi_{1}^{s}\right\rangle $ and the two vectors $\left|\varphi_{K}^{s=\pm}\right\rangle $
must be adjoined to the previous set $\left\{ \left|\psi_{i=1,2}^{s=\pm}\right\rangle \right\} $.
In this subspace, the six eigenvectors of the Hamiltonian now read
\[
\left\{ \left|\chi_{\pm}^{s=\pm}\right\rangle \equiv\frac{\left|\psi_{1}^{s}\right\rangle \pm\left|\varphi_{K}^{s}\right\rangle }{\sqrt{2}},\left|\chi_{0}^{s=\pm}\right\rangle \equiv\left|\psi_{2}^{s}\right\rangle \right\} 
\]
and the energy degeneracy is lifted. An initial state of the form
$\ket{\Psi(0)} =\sum_{k,s}c_{k}^{s}\ket{\psi_{k}^{s}}$
now has components on the six new eigenvectors, \emph{i.e.\@} 
$\ket{\Psi(0)} =\sum_{r=\pm,0}\sum_{s=\pm}d_{r}^{s}\ket{\chi_{r}^{s}}$
and therefore the time-averaged state 
\begin{align}
\label{rho}
\bar{\rho} & =\sum_{r=\pm,0}\sum_{s=\pm}\left|d_{r}^{s}\right|^{2}\left|\chi_{r}^{s}\right\rangle \left\langle \chi_{r}^{s}\right|
\end{align}
now contains a highly localized component, on the atom at position
$K$ or $\left(N-K\right)$. Accordingly, the probability distribution
$\mathcal{P}_{k}$ exhibits a strongly peaked behavior at $k=K,\left(N-K\right)$.
This localization phenomenon is in good qualitative agreement with
what we observe with the full simulation: in particular, the appearance
of the localization lines indeed happens when $\Lambda\approx\frac{7}{6}$
(see Fig.~\ref{fig:localisationExc2}). This validates the simplified
analytical treatment we have just carried out which indeed seems to
retain the main physical ingredients of the system and its evolution.

\section{Discussion}

\label{sec:Comparison}

\begin{figure}
\centering{}\includegraphics[width=8cm]{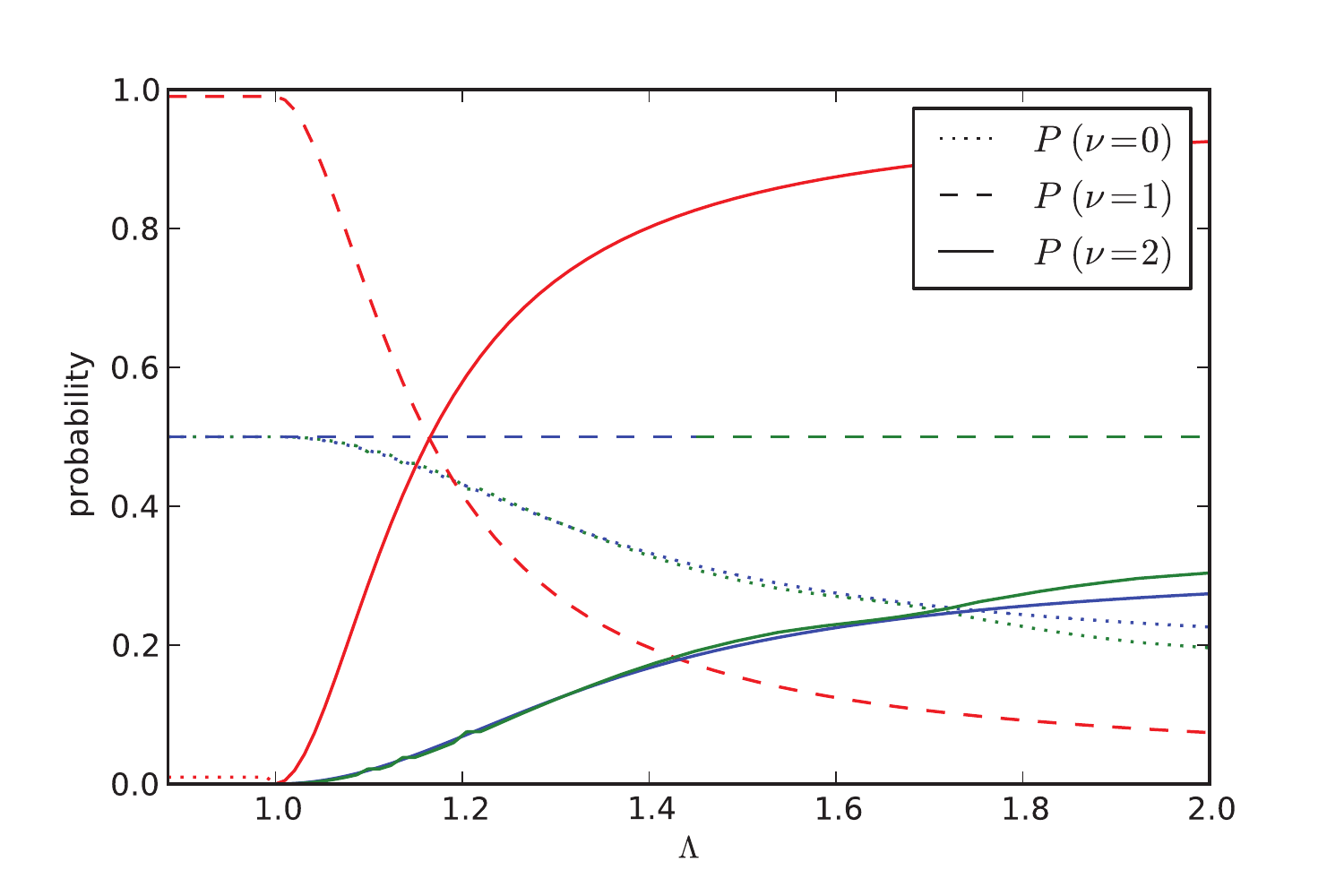}\protect\caption{\label{fig:Probabilities} Probability to have $\nu$ excitations
as a function of $\Lambda$, with $N=100$, according to the microcanonical
predictions (red), our numerical simulation (green) and the analytical
treatment (blue). }
\end{figure}

\begin{figure}
\centering{}\includegraphics[width=8cm]{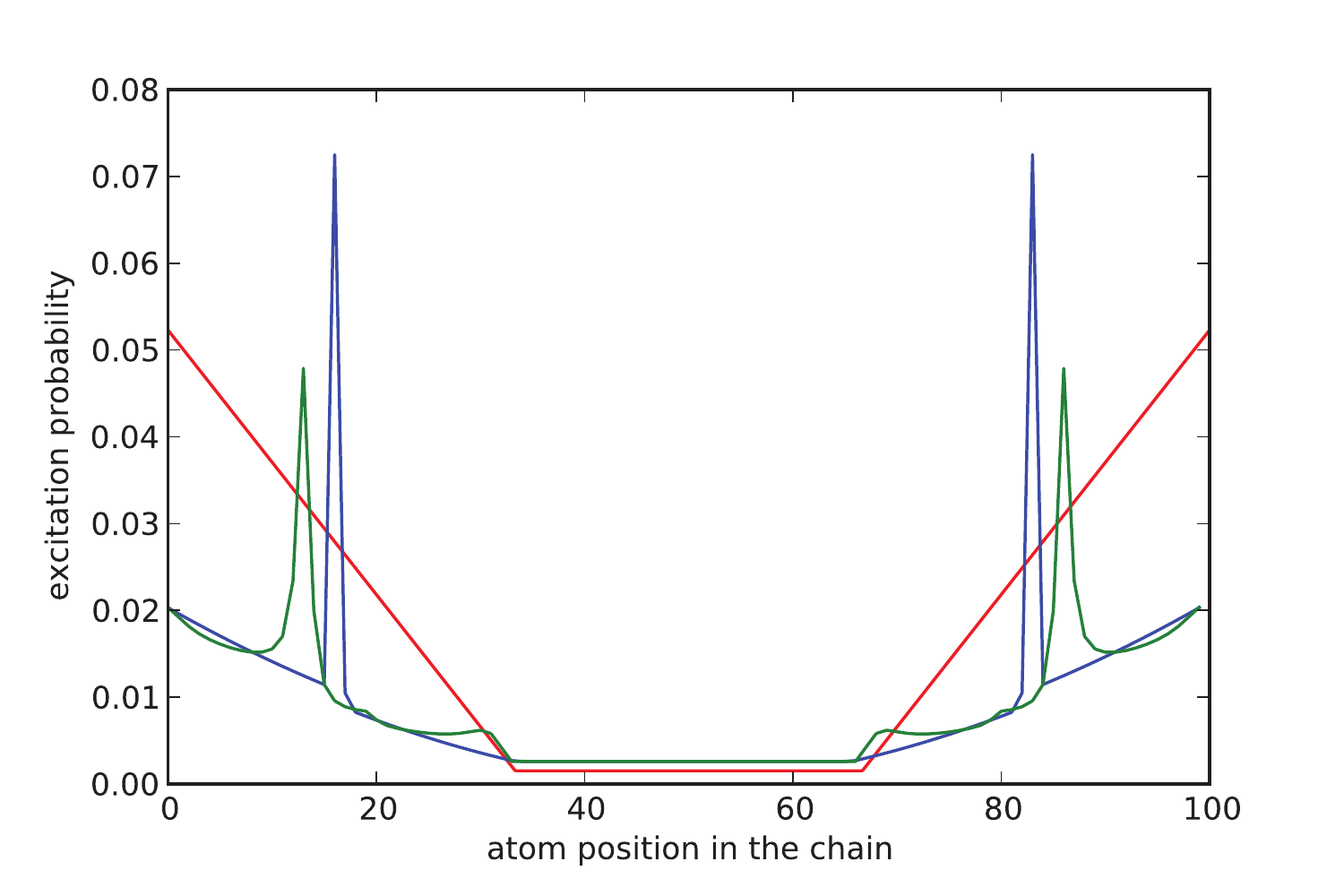} \protect\caption{\label{fig:exc_proba_1,5_} Spatial distribution of excitations at
$\Lambda=1.5$, with $N=100$, according to the microcanonical predictions
(red), our numerical simulation (green) and the analytical model (blue).}
\end{figure}

This section is devoted to the comparison of the results obtained
above following the different approaches 
and to discussions on their differences.

Fig.~\ref{fig:Probabilities} displays plots of the probability $\mathcal{P}(\nu)$
to have $\nu$ Rydberg excitations in the sample, as a function
of $\Lambda$ (for $0.9\leq\Lambda\leq2$), calculated according to:
i) the microcanonical hypothesis (Sec.~\ref{sec:Thermal}); ii)
the full simulation of the system (Sec.~\ref{sec:Numerics}); iii)
the approximate diagonalization of $H$ in a reduced $6$-dimensional
Hilbert space (Sec.~\ref{SimpleAnalyticalTreatment}). While the schemes
ii) and iii) yield very similar results (as expected), assumption
i) induces quite different behaviors. The same comparison can be performed
on the spatial probability distribution $\mathcal{P}_{k}$ which is
displayed on Fig.~\ref{fig:exc_proba_1,5_}. Again, the shapes obtained
via schemes ii) and iii) are in very good qualitative agreement: in
both cases, one observes two localization peaks on a ``background
curve'', which coincide satisfactorily. (Note that, according to
our calculations, excitations are more likely to be localized at the
borders). The spatial probability distribution obtained according
to assumption i) differs strongly: no excitation localization effect
is observed and the background curve is far from what is observed
in the full simulation.

The discrepancies observed above can be partly explained by the following
``parity balance property'' established in Appendix~\ref{App:Parity}: 
for any eigenstate $\ket\psi$ of the Hamiltonian $H$ of nonzero energy, 
the
projections $\left|\psi_{\odd}\right\rangle $ and 
$\left|\psi_{\even}\right\rangle $
onto the orthogonal and plementary subspaces $\mathcal{H}_{\odd}$
and $\mathcal{H}_{\even}$, respectively spanned by the states with
an odd and even number of Rydberg excitations, have the same norm,
\emph{i.e.\@} $\left|\psi\right\rangle =\left|\psi_{\odd}\right\rangle +\left|\psi_{\even}\right\rangle $
with $\left\Vert \left|\psi_{\odd}\right\rangle \right\Vert =\left\Vert \left|\psi_{\even}\right\rangle \right\Vert =\frac{1}{\sqrt2}$.
This property conflicts directly with the microcanonical predictions
according to which the probability to have $\nu<\nu_{\max}$ excitations
is negligible compared to the probability to have the maximum number
of excitations. For example, suppose $\nu_{\max}=1$, the microcanonical
ensemble implies that $P(\nu=0)=\frac{1}{1+N}$ and $P(\nu=1)=\frac{N}{1+N}$.
By contrast, the parity balance property implies $P(\nu=0)=P(\nu=1)=0.5$.
Furthermore, one can see that in Fig.~\ref{fig:ExcProba}, each time
one of the probability curve is above $\frac{1}{2}$, the parity balance
condition is therefore impossible to fulfill. In almost all cases,
the even/odd parity balance property and the simple microcanonical
approach presented in Sec.~\ref{sec:Thermal} disagree.

The inaccuracy of the predictions deduced from the microcanonical
assumption can also be explained by the choice of 
$\left|\varnothing\right\rangle $
as initial state: the low connectivity of this state to the rest
of the Hilbert space constitutes indeed a strongly limiting factor
to the thermalization process \cite{OML10}. In particular, the vaccum
state being symmetric as well as the Hamiltonian, the system remains
in a symmetric state during its evolution. The direct application
of the microcanonical assumption, taking into account all the states
which are allowed by the Rydberg blockade, is therefore incorrect
: for a proper use of the microcanonical hypothesis, one should actually
take this extra symmetry selection rule into consideration and count
only the accessible, \emph{i.e.\@} symmetric, states. Note that the
vacuum state is the natural starting point from an experimental perspective
to study the build-up of excitations and is therefore widely used
\cite{AL12,BMFALA13}.

Another choice of initial state can actually be considered. Starting
with a random initial state, Ates et al. \cite{AGL12} showed that
in the regime of strong nearest neighbor interaction ($\Lambda>\frac{N}{2}$),
the dynamics of the system is well described by the microcanonical
ensemble. In the regime studied in the present article, $\Lambda\ll N$,
a similar random choice of initial state leads to an essentially ``frozen
evolution'' as seen by the following dimensionality arguments. From
Eq.~\eqref{eq:Nnu2}, the number of states containing at most $\nu_{\max}-1$
excitations is $\propto N^{\nu_{\max}-1}$ and the dimension of the
generated subspace $\mathcal{H}_{\nu\leq\nu_{\max}-1}$ is a small
fraction $O(\frac{1}{N})$ of the dimension of the total Hilbert space
$\mathcal{H}$. As $N$ increases, the Hilbert space is therefore
essentially composed by states containing $\nu_{\max}$ excitations.
Furthermore, since all eigenvectors of $H$ with non-zero eigenvalue
follow the parity balance property, 
\begin{align*}
\text{dim}(\mathcal{H}\setminus\text{ker}(H)) & \leq2\min\left(\text{dim}(\mathcal{H}_{\text{even}}),\text{dim}(\mathcal{H}_{\text{odd}})\right)\\
 & \sim2\text{dim}\left(\mathcal{H}_{\nu=\nu_{\max}-1}\right)\\
 & \sim O\left(N^{\nu_{\max}-1}\right)
\end{align*}
As a consequence, the Hilbert space is mainly spanned by the states
in $\mathrm{ker}\left(H\right)$ with $\nu_{\max}$ excitations. 
This can be seen, for example, on Fig.~\ref{fig:DiffEnergy}. 
Therefore,
the projector on $\text{ker}(H)$ is a ``gentle'' operator \cite{W99} for the
ensemble of states picked uniformly at random: with high probability,
a state from this ensemble will have a large component on $\text{ker}(H)$
and its evolution will essentially be ``frozen'', which contradicts
the microcanonical predictions. 

Conversely, if one chooses the initial
state in the $\mathcal{H}_{\nu\leq\nu_{\max}-1}$ subspace, the system
will not explore $\text{ker}(H)$: the dimensionality of the actual
microcanonical ensemble is therefore again much less than the number
of states allowed by the Rydberg blockade.
Note that this initial state choice is a natural generalization of 
$\ket{\varnothing}$ to study the buildup of excitations.

\section{Conclusion}

\label{sec:Conclusion}

In the present article, we studied the dynamics of a 1D-Rydberg ensemble
in the regime of at most 2 excitations. In the same conditions as
in \cite{AL12,BMFALA13}, we tested the validity of the microcanonical
predictions and found it cannot be used straightforwardly to account
for the thermalization process which occurs in this particular regime.
Though the observed discrepancies can be related to our specific choice
of initial state and its particular symmetry properties, we also proved,
by an argument involving the dimension of the kernel of the Hamiltonian,
that the same restriction holds for a randomly chosen initial state.

Further investigations are needed to better understand when and how
to apply the (micro)canonical predictions. In particular, the results
presented here all rely on the hardcore sphere assumption. Refining
the model and considering the full Rydberg-Rydberg interaction Hamiltonian
Eq.~\eqref{V_dd} might actually change our conclusions and make
the microcanonical assumption more adapted, as shown in \cite{LOG10}.
Indeed, in that case, all states become, strictly speaking, allowed,
though more or less accessible, and the connectivity accordingly increases
between states of the Hilbert space. Moreover, as suggested by our
discussion, the systematic study of symmetry properties of the system
at stake, as well as the selection rules they impose, appears to be
a crucial point in the proper application of microcanonical assumption.

\section*{Acknowledgments}
We thank Stefano Bettelli for digging up the raw data \cite{BMFALA13data}.
Maurice Raoult's highlighting of the simplicity of 
$2\times2$ matrix diagonalization inspired us for
the simplified analytical treatment of section \ref{SimpleAnalyticalTreatment}.

\appendix

%\label{sec:Appendix}
\section{Quantitative agreement of our analytical treatment and Bettelli 
  \emph{et al.\@}'s Monte-Carlo Simulation}
\label{App:Comparison}

In this appendix, we compare our analytical treatment of the microcanonical
hypothesis with the Monte-Carlo results published by Bettelli \emph{et al.\@} 
in \cite{BMFALA13}.

On their Fig.~2, Bettelli \emph{et al.\@}
give the average number of excitation $\braket{\nu}$
and its standard deviation $\sqrt{\braket{\Delta\nu^2}}$ for %three values of
$\Lambda\in\{2.1, 2.45, 3.15\}$. 
These data points correspond to the crosses on
Fig.~\ref{fig:AvrgNu} and fall on the corresponding curves computed
according to our analytical treatment.
For 5 of these 6 values, our results are indeed 
identical to the two published decimals. 
The 6th value is the standard deviation for $\Lambda=3.15$, 
where we obtain $\sqrt{\braket{\Delta\nu^2}}=0.41$, to be compared to $0.38$. 
This deviation is small, and we therefore consider the results to be 
effectively identical.

\begin{figure}
 \centering{}\includegraphics[width=8cm]{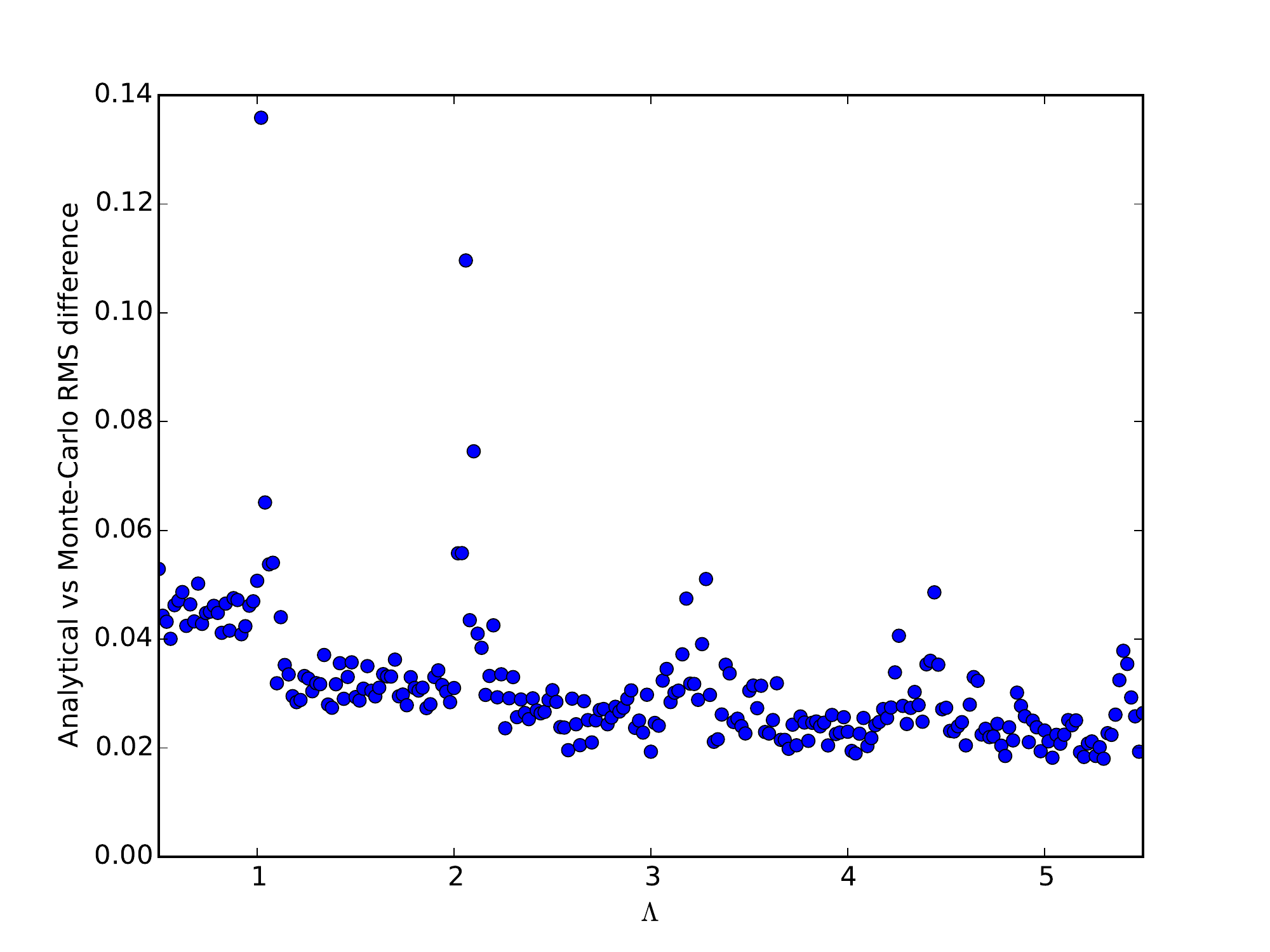}%
 \protect\caption{\label{fig:rmsdiff} Root-mean-square difference between the 
  Monte-Carlo result of \cite{BMFALA13} and the analytical 
  Eq.~\eqref{eq:excdensity}
  as a function of $\Lambda$.}
\end{figure}
\begin{figure}
 \centering{}\includegraphics[width=8cm]{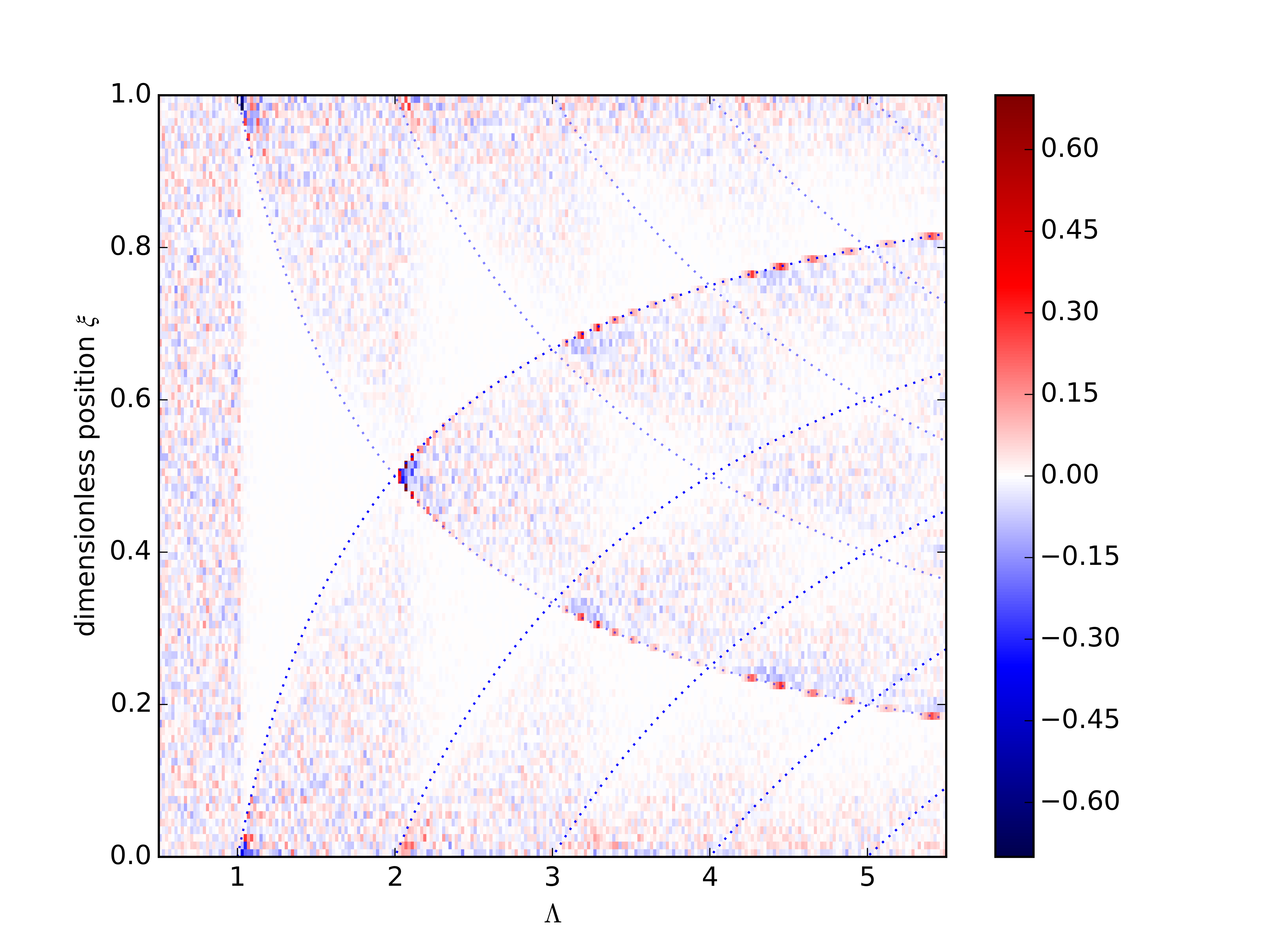}%
 \protect\caption{\label{fig:diff} Difference between the Monte-Carlo 
 result of \cite{BMFALA13} and the analytical result Eq.~\eqref{eq:excdensity}.
 On the dotted lines,
 either $\xi$ or $1-\xi$ is an integer multiple of $\frac1{\Lambda}$.}
\end{figure}

We compared
the spatial distribution of excitations of Fig.~\ref{fig:ExcDistribution}
to the data \cite{BMFALA13data} kindly provided bay the authors of
\cite{BMFALA13}. 
This data-set was obtained with a Monte-Carlo simulation with $N=10^4$ atoms and
$N_{\text{rep.\@}}=5\times10^4$ repetitions, 
using $N_{\text{bin}}=100$ bins and a normalization to an average excitation density of 1.

We plotted the Monte-Carlo simulation and our data, 
computed from Eq.~\eqref{eq:excdensity}, using the same normalization
and we were unable to visually see any difference by blinking between the two 
plots on our computer screen. 
More quantitatively, we plotted the root-mean-square difference between the 
two sets of data for each
value of $\Lambda$ on Fig.~\ref{fig:rmsdiff}, as well as the pixel by pixel
difference  on Fig.~\ref{fig:diff}.
  
When $\Lambda<1$, the
probability to have one excitation in any given bin is
$\frac1{N_{\text{bin}}}$; the expected value of this root-mean-square difference
as well as the standard deviation of the difference should then be 
$\sqrt{\frac{N_{\text{bin}}}{N_{\text{rep}}}}=\sqrt{\frac1{500}}\simeq 0.045$. 
When $\Lambda\geq1$, no strong localization is expected, and this calculation 
should therefore give a correct order of magnitude, 
both for the root-mean-square difference for a given $\Lambda$ and 
for the pixel by pixel fluctuations. 
This is quantitatively consistent with the results.

Furthermore, the main deviations in both graphs can be explained by the different 
approximations in plotting each pixel : for the Monte-Carlo simulation 
\cite{BMFALA13data}, 
the value of a pixel of coordinates $(\Lambda, \xi)$ 
corresponds do an average over the
segment $\left[\xi,\xi+\frac{1}{N_{\text{bin}}}\right)$, 
while, for the analytical formula \eqref{eq:excdensity}, we
computed its value at the center of the pixel, \emph{i.e.\@} for  
$\xi+\frac{1}{2N_{\text{bin}}}$. 
The latter approximation, taken for the sake of simplicity, is only justified
when Eq.~\eqref{eq:excdensity} is reasonably flat.
The main deviations seem indeed to be localized where the latter
approximation is not justified, \emph{i.e.\@} 
when the excitations are concentrated in a few narrow peaks,
or where either $\xi$ or $1-\xi$ is an integer multiple of $\frac{1}{\Lambda}$.

Our analytical treatment of the microcanonical ensemble assumption is therefore
quantitatively consistent with the Monte-Carlo simulation in \cite{BMFALA13,
BMFALA13data}.

\section{Parity balance property}
\label{App:Parity}

Our analytical model has been build according to an observation on the structure of $H$: the parity balance property presented in Sec.~\ref{sec:Comparison}.
The Hilbert space $\mathcal{H}$ can be decomposed into 2 orthogonal subspaces containing an even/odd number of excitations: $\mathcal{H}= \mathcal{H}_{\even}  \oplus \mathcal{H}_{\odd}$. 
Since $H$ either removes or adds an excitation, its effect on a state containing
an even (resp. odd) number of excitations will change the parity of its number 
of excitations to an odd (resp. even) value. From this, we can deduce that
the subspaces $\mathcal{H}_{\even}$ and $\mathcal{H}_{\odd}$ are stable under the application of $H^2$. 
$H^2$ is therefore in the form: 
\begin{equation}
H^2=
\begin{pmatrix}
H^2_{\even} & 0 \\
0 & H^2_{\odd} \\
\end{pmatrix}
\label{eqHH}
\end{equation}

Our analytical treatment is based on the diagonalization of $H^2_{\even}$ and $H^2_{\odd}$ to obtain the eigenstates of $H$. 
In particular, the eigenstates of $H$ follow this even/odd decomposition and can be written as $\Ket{\Psi}=\Ket{\Psi_{\even}}+\Ket{\Psi_{\odd}}$ 
with $\Ket{\Psi_{\even/\odd}}=\Pi_{\even/\odd}\Ket{\Psi}$ with $\Pi_{\even/\odd}$ being the projector on $\mathcal{H}_{\even/\odd}$. 
Using the orthogonality of $\Ket{\Psi_{\even}}$ and $\Ket{\Psi_{\odd}}$, a short calculation from $H \Ket{\Psi_{\even/\odd}}= E \Ket{\Psi_{\odd/\even}}$ leads to either $E = 0$, or $\norme{\Ket{\Psi_{\even}}}= \norme{\Ket{\Psi_{\odd}}}  =\frac{1}{\sqrt{2}}$,
i.e. all non zero energy eigenstates are equally weighted between even and odd parts. 

If $\Lambda<2$, it is straightforward to see that all $E=0$ eigenstates are 
orthogonal to $\Ket{\varnothing}$. If the initial state is $\Ket{\varnothing}$, 
the average probability to have an even (or odd) number of excitations is therefore
exactly $\frac{1}{2}$.

\section{Diagonalization of $H^2$}
\label{App:DiagH2}

Here, we present a more detailed version of the diagonalization of $H^2$ showed 
in Sec.~\ref{SimpleAnalyticalTreatment} using its even/odd decomposition 
described by Eq.~\eqref{eqHH} in Appendix~\ref{App:Parity}. 
The states defined by 
Eqs.~\eqref{state1},\eqref{state2} and \eqref{state3} have the following explicit formulations:
\begin{align*}
\left|\phi_{1}\right\rangle &= \frac{1}{\sqrt{N}}\sum_{k=0}^{N-1} \Ket{k}\\
\left|\phi_{2}\right\rangle &= Z_{2} \sum_{k=0}^{N-n_b-1} { \sum_{l=k+n_b}^{N-1} { 2\Ket{k,l}}}\\
\left|\phi_{3}\right\rangle &= Z_{3} \left(\sum_{k=0}^{N-n_b-1} { (N-n_b-k) \Ket{k} }  + \sum_{l=n_b}^{N-1} { (l-n_b) \Ket{l} } \right)
\end{align*}
When $N\gg 1$, the sums involved in the computation of $\Braket{\phi_{2}|\phi_{2}}$ and $\Braket{\phi_{3}|\phi_{3}}$ can easily be approximated by integrals, and the normalization factors
are therefore:
\begin{align}
Z_2 &\simeq \frac{\sqrt{3/2}}{(N-n_b)^{3/2}} \\
Z_3 &\simeq \frac{1}{\sqrt{2}(N-n_b)}
\end{align}

$H^2_{\even}$ can be expressed in the basis $\{\Ket{\varnothing}, \Ket{\phi_{2}}  \}$:
$$H^2_{\even}=\Omega^2 N
\begin{pmatrix}
1 & \rho \sqrt{2} \\
\rho\sqrt{2} & \frac{4}{3}\rho \\
\end{pmatrix}$$
with $\rho = \frac{N-n_b}{N} \simeq 1-\frac{1}{\Lambda}$. The eigenvalues are
\begin{align}
  \label{eq:E2pmeven}
  E^2_{\even,1} &= \Omega^2 N \frac{4\rho +3 - \sqrt{ 88\rho^{2} - 24\rho + 9}}{6}\\
  E^2_{\even,2} &= \Omega^2 N \frac{4\rho +3 + \sqrt{ 88\rho^{2} - 24\rho + 9}}{6}
\end{align}
for respectively $\Ket{E^2_{\even,1}}$ and $\Ket{E^2_{\even,2}}$.

$H^2_{\odd}$ can be expressed in the orthonormal basis $\{\Ket{\phi_{1}}, \Ket{\phi_1^\perp}= Z_1^{\perp}(\Ket{\phi_3} - \Braket{\phi_1|\phi_3} \Ket{\phi_{1}} )\}$ with $Z_1^{\perp}=\frac{1}{\sqrt{1-\frac{3\rho}{2}}}$.
$$H^2_{\odd}=\Omega^2 N
\begin{pmatrix}
1+2\rho^2 & 2\rho^{3/2}\sqrt{\frac{2}{3}-\rho} \\
2\rho^{3/2}\sqrt{\frac{2}{3}-\rho} & \frac{4}{3}\rho - 2\rho^2 \\
\end{pmatrix}$$
Diagonalizing $H^2_{\odd}$ in the same way, we find its eigenvalues $E^2_{\odd,i} = E^2_{\even,i}=E^2_{i}$ corresponding to the eigenvectors $\Ket{E^2_{\odd,i}}$ with $i=1,2$.

Each subspace with eigenvalue $E^2_{i=1,2}$ has dimension 2 and 
is thus generated by 
$\frac{ \Ket{E^2_{\even,i}} + e^{i\phi}\Ket{E^2_{\odd,i}} }{\sqrt{2}}$ and
$\frac{ \Ket{E^2_{\even,i}} - e^{i\phi}\Ket{E^2_{\odd,i}} }{\sqrt{2}}$. Let us fix the arbitrary relative phase $\phi$ to 0 by the equation:
$$H\Ket{E^2_{\even,i}}= \sqrt{E^2_{i}}\Ket{E^2_{\odd,i}}.$$
Combining it with $H^2\Ket{E^2_{\even,i}}=E^2_{i}\Ket{E^2_{\even,i}}$  gives
$$H\Ket{E^2_{\odd,i}}= \sqrt{E^2_{i}}\Ket{E^2_{\even,i}}.$$

We trivially define the 4 eigenvalues of $H$ by $E^{s=\pm}_{i=1,2}= s \times E_{i=1,2} = s \times \sqrt{E^2_{i=1,2}}$ of the 4 eigenvectors 
$\ket{\psi^{\pm}_{i=1,2}}  = \frac{ \ket{E^2_{\text{even},i}} + s \times \Ket{E^2_{\odd,i}} }{\sqrt{2}}$.

\section{Localization of excitations} 
\label{App:Localization}

Now, we present specifically our analytical treatment of the localization effects observed in Fig.~\ref{fig:localisationExc2}.
The states $\Ket{\phi_{1}}$, $\Ket{\phi_{2}}$ and $\Ket{\phi_{3}}$ used in Sec.~\ref{SimpleAnalyticalTreatment} are delocalized. To account for the very narrow
localization of excitations (1 atom wide), we complete the four states 
$\ket{\psi^{\pm}_{i=1,2}}$ 
with the family of states 
$\ket{\varphi_{k}^{s=\pm}}$ defined in Eq.~\eqref{state}.

For large $N$, those states are approximate eigenstates of $H$ with eigenvalues $s \times \varepsilon_k= s \times \Omega\sqrt{N-n_b-k} $:
\begin{equation}
H\Ket{\varphi_{k}^{s=\pm}}= 
  s \times \Omega\sqrt{N-n_b-k} \Ket{\varphi_{k}^{s=\pm}} 
  + O\left(\frac{1}{\sqrt{N}}\right). 
\end{equation}

As seen in section \ref{SimpleAnalyticalTreatment}, for a particular value $K$,
the state $\Ket{\varphi_{K}^{s=\pm}}$ has the same energy as the collective
excitation state $\Ket{\psi^s_1}$. 
This resonance can only exist if 
\begin{equation}
E_1 \leq \varepsilon_0 \Leftrightarrow %
	\Omega \sqrt{N \frac{4\rho +3 - \sqrt{ 88\rho^{2} - 24\rho + 9}}{6}} %
		\leq \Omega \sqrt{N \rho}
\end{equation} 
This inequality holds when $\Lambda \geq \frac{7}{6}$. 
 
The small coupling between those states lifts this degeneracy by adding an energy shift $\pm \delta$ to the new eigenvectors 
	$\Ket{\chi^{s=\pm}_{\pm}}=%
	\frac{\Ket{\varphi_{K}^{s}} \pm \Ket{\psi^s_1}}{\sqrt{2}}$
	of energy $s\times E_1 \pm \delta$.
This degeneracy allows us to keep only the diagonal terms in the density matrix
after time averaging. 
Furthermore, when $\ket{\Psi(0)}$ is reasonably delocalized 
state, like \emph{e.g.\@} $\ket{\varnothing}$,
we have 
$\Braket{\chi^s_\pm|\Psi(0)}=%
	\frac{ \Braket{ \psi^s_1|\Psi(0)} }{\sqrt{2}} + O(\frac{1}{\sqrt{N}})$. 
After time averaging, the density matrix contains only diagonal
terms and one can deduce the density matrix:
\begin{widetext}
\begin{equation}
\bar\rho = 
	\left|\Braket{ \psi^+_2|\Psi(0)} \right|^2 
		\left(\Ket{\psi^+_2}\Bra{\psi^+_2} + \Ket{\psi^-_2}\Bra{\psi^-_2}\right) 
  + \frac{|\Braket{ \psi^-_1|\Psi(0)} |^2}{2} 
		\left( \Ket{\psi^-_1}\Bra{\psi^-_1} + \Ket{\psi^+_1}\Bra{\psi^+_1}  %\\ 
    + \Ket{\varphi_K^{+}}\Bra{\varphi_K^{+}} 
		+ \Ket{\varphi_K^{-}}\Bra{\varphi_K^{-}}\right)
\end{equation}
\end{widetext}

The average state $\bar\rho$ of the system is therefore a statistical mixture involving 
$\Ket{\psi^+_2}$ and $\Ket{\psi^-_2}$ weighted by 
$\left|\Braket{ \psi^+_2|\Psi(0)} \right|^2$ 
and the four states $\Ket{\psi^-_1}$, $\Ket{\psi^+_1}$,$\Ket{\varphi_K^{-}}$
and $\Ket{\varphi_K^{+}}$ weighted by 
$\tfrac12{\left|\Braket{\psi^-_1|\Psi(0)}\right|^2}$. 
This is the explicit form of the density matrix described by Eq.~\eqref{rho}.

\end{document}